\documentclass[twocolumn,amsmath,amssymb,superscriptaddress,prl]{revtex4}

\usepackage{graphicx}
\usepackage{amsmath,amssymb,amsthm}

\newcommand{\beq}{\begin{equation}}
\newcommand{\eeq}{\end{equation}}
\newcommand{\beqa}{\begin{eqnarray}}
\newcommand{\eeqa}{\end{eqnarray}}

\newcommand{\w}{\omega}

\newcommand{\ket}[1]{\left| #1 \right\rangle}

\begin{document}

\title{Comment on ``General Non-Markovian Dynamics of Open Quantum System''}
\author{Dara P. S. McCutcheon}
\affiliation{\mbox{Departamento de F\'isica, FCEyN, UBA and IFIBA, Conicet, Pabell\'on 1, Ciudad Universitaria, 1428 Buenos Aires, Argentina}}
\author{Juan Pablo Paz}
\affiliation{\mbox{Departamento de F\'isica, FCEyN, UBA and IFIBA, Conicet, Pabell\'on 1, Ciudad Universitaria, 1428 Buenos Aires, Argentina}}
\author{Augusto J. Roncaglia}
\affiliation{\mbox{Departamento de F\'isica, FCEyN, UBA and IFIBA, Conicet, Pabell\'on 1, Ciudad Universitaria, 1428 Buenos Aires, Argentina}}

\date{\today}

\maketitle

The existence of a ``non-Markovian dissipationless'' regime, 
characterized by long lived oscillations, was reported in Ref.~\cite{zhang12} for a class of quantum open systems.
It is claimed this could happen in the strong coupling 
regime, a surprising result which has attracted some attention. 
We show that this regime exists if and only if the total  
Hamiltonian is unbounded from below, 
casting serious doubts on the usefulness of this result. 

We focus on the simplest bosonic model discussed 
in~\cite{zhang12}: An oscillator  
$H_S=\Omega\, a^\dagger a$ couples to an environment  
$H_E=\sum_k \omega_k b^\dagger_k b_k $ 
through the interaction  $H_{I}=\sum_k\lambda_k(a b^\dagger_k + a^\dagger b_k)$. 
The total Hamiltonian $H_T=H_S+H_E+H_{I}$ commutes with the excitation number   
$N=a^\dagger a +\sum_k b^\dagger_k b_k$. 
Thus, an eigenstate of $H_T$ in the single excitation sector can be written  
$\ket{\Phi}=C^{\dagger}\ket{0,0}$, with 
$C^{\dagger}=c_s a^\dagger + \sum_k c_k b^\dagger_k$ a generalized creation operator 
and $\ket{0,0}$ the vacuum. 
Requiring $H_T\ket{\Phi}=E\ket{\Phi}$ implies
$Ec_s=\Omega c_s+\sum_k\lambda_k c_k$ and 
$Ec_k=\omega_k c_k+\lambda_k c_s$~\cite{rancon14}. The energy $E$ therefore satisfies 
$E=\Omega+\sum_k \lambda^2_k/(E-\omega_k)$, which for negative solutions $E=-|\omega_0|$ becomes 
$\Omega+|\omega_0|=
\sum_k \lambda^2_k/(|\omega_0|+\omega_k)$. This equation has solutions if and only if 
$\Omega+\delta\Omega<0$ with 
$\delta\Omega=-\sum_k\lambda^2_k/\omega_k$. 
Thus, in this regime the {\emph{total}} Hamiltonian acquires a negative eigenvalue. Moreover, 
noticing that $[H_T,C^{\dagger}]=E C^{\dagger}$, we see that there also exist eigenstates $\ket{\Phi_n}=(C^{\dagger})^n\ket{0,0}$ with 
eigenvalues $nE=-n|\w_0|$, extending to negative infinity: the total Hamiltonian is unbounded from below in this regime.
Unbounded Hamiltonians, such as $H=-\w a^{\dagger} a$ 
have no ground state, no thermal state (divergent partition functions), and would act as infinite 
sources of energy when weakly coupled with any other 
system (they are thermodynamically unstable).

We now follow the argument presented 
in~\cite{zhang12} to show that the dissipationless 
regime at strong coupling is precisely the regime 
when $\Omega+\delta\Omega<0$, and the total 
Hamiltonian is unbounded. 
The density matrix of the system 
satisfies~\cite{zhang12,RoncagliaPRA},
$\dot{\rho}=-i[\tilde\Omega(t) a^\dagger a,\rho]+
\gamma(t)(1+\tilde n(t))(2 a\rho a^\dagger-a^\dagger a\rho-
\rho a^\dagger a)
+\gamma(t)\tilde{n}(t)(2 a^\dagger \rho a-a a^\dagger\rho-\rho a a^\dagger ).$ 
Coefficients depend on the 
Green's function which satisfies 
$\dot u(t)+i \Omega u(t) +\int^t_0 \mathrm{d}s\, \eta(t-s) u(s)=0$, 
where the dissipation kernel is 
$\eta(s)=\int_0^\infty \mathrm{d}\omega J(\omega) \exp[-i\omega s]$, 
and spectral density
$J(\omega)=\sum_k\lambda_k^2\delta(\omega-\omega_k)$. 
The frequency and damping rate satisfy  
$i \tilde{\Omega}(t)+\gamma(t)=-\dot{u}(t)/ u(t)$ 
while $\gamma(t)\tilde{n}(t)=\frac{1}{2}\dot{\xi}(t)+\gamma(t)\xi(t)$, where 
$\xi(t)=\int_0^t\mathrm{d}\tau\int_0^t\mathrm{d}s\, u(\tau) \tilde{\nu}(s-\tau)u(s)^*$ and  
$\tilde{\nu}(s)=\int_0^\infty \mathrm{d}\omega J(\omega) 
\exp[i\omega s]/(\exp(\omega/k_BT)-1)$ ($T$ is the 
temperature of the environment). This equation is 
valid for all spectral densities and 
temperatures, and is the tool used to study
non--perturbative and non--Markovian effects. 

\begin{figure}
\begin{center}
\includegraphics[width=0.45\textwidth]{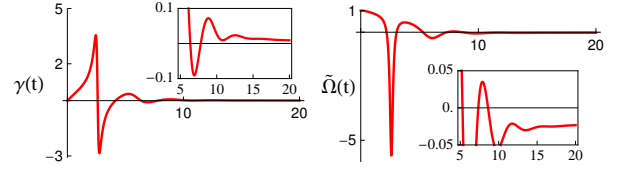}
\caption{Damping rate and renormalized 
frequency for the dissipationless sub--ohmic model shown
in Fig.~(2) of Ref.~\cite{zhang12}.}
\label{plots}
\vspace{-0.5cm}
\end{center}
\end{figure}

A dissipationless regime exists when $u(t)\to r \exp[-i\w_0 t]$ at long times. 
In this case, $\Omega(t)=\w_0$, $\gamma(t)=\gamma(t)\tilde n(t)=0$: as $t\to \infty$ the system evolves unitarily with a 
Hamiltonian $\tilde H_S=\w_0\,a^\dagger a$. 
For $u(t)$ to behave in this way, its 
Laplace transform must have a purely imaginary pole, i.e. 
$\omega_0- \Omega  +i \hat\eta(-i\omega_0)=0$, 
where the Laplace transform of $\eta(t)$  is 
$\hat{\eta}(s)$. The imaginary part of this 
equation is $J(\omega_0)=0$. For spectral densities of any type (Ohmic, sub-Ohmic etc.) satisfying $J(\w)>0$ for all $\w>0$, this condition 
can be satisfied for $\w_0<0$~\cite{zhang12}. 
With $\w_0=-|\w_0|$ the real part gives $\Omega+|\omega_0|= 
\int_0^\infty \mathrm{d}\omega J(\omega)/(\omega+|\omega_0|)$, which has solutions 
if and only if $\Omega+\delta\Omega<0$, with
$\delta\Omega=-\int_0^\infty \mathrm{d}\w J(\omega)/\omega$. This is precisely the condition under which 
the total Hamiltonian becomes unbounded.
This also manifests in the 
master equation. In Fig.~({\ref{plots}}) we plot 
$\gamma(t)$ and $\tilde\Omega(t)$ for the same
parameters as in Fig.~(2) of Ref.~\cite{zhang12} 
(where $\Omega(t)$ was not analyzed). 
This exact calculation shows 
that $\tilde\Omega(t)$ approaches a negative value while
the damping rate vanishes, making the renormalized
Hamiltonian also unbounded. 

Thus, we have proved that the dissipationless regime of 
Ref.~\cite{zhang12} for strong coupling exists if 
and only if the total Hamiltonian is unbounded from 
below, and therefore thermodynamically unstable. 
An analogous instability is well 
known for the famous model where 
the system Hamiltonian is $H_S=p^2/2m+\kappa x^2/2$, 
the environment is 
$H_E=\sum_k(p^2_k/2m_k+m_k\omega^2_k q^2_k/2)$
and the interaction is 
$H_{I}=\sum_k c_k x q_k$. 
A simple calculation shows
that the total Hamiltonian is $H_T=H_R+
\sum_k \{ p_k^2/2m_k+(m_k \omega_k^2 
q_k+c_k x)^2/2m_k\omega^2_k\}$. 
Here, $H_R=p^2/2m+\kappa_Rx^2/2$ with 
$\kappa_R= \kappa+\delta\kappa$ and 
$\delta\kappa=-\sum_k\lambda^2_k/m_k\omega^2_k$~\cite{Martinez13}. 
Thus, when $\kappa_R<0$, the total Hamiltonian $H_T$ is unbounded (and, as opposed to the previous 
case, is both thermodynamically and 
dynamically unstable). 

In Ref.~\cite{zhang12} the authors identify long lived 
oscillations when $J(\omega)$ has band gaps. We do 
not question this result, which is indeed valid but 
far from surprising.

\end{document}